\documentclass[11pt,fleqn]{article}
\usepackage{graphics}	
\usepackage{latexsym}
\usepackage{bm}

\begin{document}
\baselineskip = 4.5mm

\begin{center}

{\bf \Large T-matrix of $Z^0$ Decay into Two Photons }

\vspace{1cm}

N. Kanda\footnote{nkanda@phys.cst.nihon-u.ac.jp}, 
R. Abe\footnote{rabe@phys.cst.nihon-u.ac.jp}, 
T. Fujita\footnote{fffujita@phys.cst.nihon-u.ac.jp}, 
H. Kato\footnote{hhkato@phys.cst.nihon-u.ac.jp}, 
and H. Tsuda\footnote{nobita@phys.cst.nihon-u.ac.jp}

Department of Physics, Faculty of Science and Technology, 

Nihon University, Tokyo, Japan

\vspace{2cm}

{\bf Abstract}

\end{center}

We first confirm that Nishijima's method of the $\pi^0 \rightarrow 
\gamma +\gamma $ calculation can precisely reproduce the observed life time 
of $\pi^0$ decay. Then, we calculate, for the first time, the T-matrix 
of the $Z^0 \rightarrow \gamma +\gamma $ process in which  the vertex of 
the $\gamma^\mu \gamma_5$ is responsible for the decay of the weak vector 
boson.  Even though the decay rate itself vanishes to zero due to 
the symmetry nature of two photons (Landau-Yang theorem), the T-matrix of 
the process has neither linear nor logarithmic divergences. Therefore, 
there is no room for the regularization of the triangle diagrams with 
the $\gamma^\mu\gamma^5$ vertex. Further, the T-matrices of all the triangle 
diagrams do not have any divergences at all, and therefore it is rigorously 
proved that the anomaly equation is spurious and appears only because of 
the improper regularization of unphysical amplitudes.

%\pacs{14.70.Hp,11.10.Gh,11.15.-q}

\section{Introduction}

For a long time, people believe that the decay rate of $\pi^0 \rightarrow 2 \gamma$ 
should be described in terms of the anomaly equation which is proposed by Adler 
in 1969 \cite{adler}. This is somewhat surprising since the decay rate of 
$\pi^0 \rightarrow 2 \gamma$ is well described by the calculation of the normal 
Feynman diagrams before the work of Adler. In fact, Nishijima explained 
the pion decay process quite in detail in the field theory textbook in 1969 
\cite{nishi}, and presented his calculation of the decay process of $\pi^0 
\rightarrow 2 \gamma$ in terms of the standard perturbation calculation. 
The calculated result of the decay width is given as
$$ \Gamma_{\pi^0\rightarrow 2\gamma}={\alpha^2\over 16\pi^2}{g_{\pi}^2\over 4\pi} 
\left( {m_{\pi}\over M_N} \right)^2 m_{\pi} \eqno{(1.1)} $$
where $\alpha$, $m_{\pi}$ and $ M_N $ denote the fine structure constant, the mass of 
pion and the mass of nucleon, respectively. Therefore, there is no anomaly 
in the T-matrix evaluation of $\pi^0 \rightarrow 2 \gamma$ if one carries out 
the calculation properly. Here, there is no ambiguity of the T-matrix 
evaluation since all of the divergences vanish to zero at the level of Trace 
evaluations. 

However, Adler claimed in 1969 that the T-matrix of the two photon decay with 
the axial vector coupling $\gamma^\mu \gamma^5$ should be regularized because of 
the linear divergence, and then he derived the anomaly equation. 
In his paper, however, one sees that he did not include the Feynman diagram 
of two photons interchanged, and therefore his calculation is not connected 
to a physical triangle diagram. As seen below, the correct T-matrix of 
the $Z^0 \rightarrow 2 \gamma $ process does not have either linear nor logarithmic 
divergences, and the linear divergence Adler found is due to the improper triangle 
diagram with $\gamma^\mu \gamma^5$ interaction. 

In this paper, we first confirm that the decay rate of 
$\pi^0 \rightarrow 2 \gamma$ can be indeed described by Nishijima's calculation.
The interaction between pions and nucleons can be described in terms of 
the pseudoscalar coupling as 
$$ {\cal L}_I =ig_\pi \bar{\psi} \gamma_5 \bm{\tau} \psi \cdot \bm{\varphi} 
\eqno{(1.2)} $$ 
where $g_\pi $ denotes the pion-nucleon coupling constant. If we take the value of 
the pion-nucleon coupling constant $g_\pi$ as ${g_\pi^2\over 4\pi }\simeq 8 $ 
which is slightly smaller than the value as suggested from the nucleon-nucleon 
scattering data \cite{bohr}, we obtain 
the decay width of $ \Gamma_{\pi^0\rightarrow 2\gamma} \simeq 7.5 \ \ {\rm eV} $
which should be compared with the observed value of 7.8 eV. Note that the $\pi N N$ 
coupling constant in the pion decay should include the form factor effect of the $N-N$ 
scattering case, and thus it should be smaller than the scattering value. 

In the same way, we can calculate the decay rate of the $Z^0 \rightarrow 2 \gamma $ 
process. In this case, the interaction Lagrangian density for the $Z^0$ boson $Z^\mu $ and 
fermions $\psi_\ell$ can be written as \cite{mandl}
$$ {\cal L}_{II} =g_z \bar{\psi_\ell}\gamma^\mu \gamma_5\psi_\ell  Z_\mu 
-0.06 g_z \bar{\psi_\ell}\gamma^\mu\psi_\ell  Z_\mu \eqno{(1.3)} $$ 
which is obtained from the standard model weak Hamiltonian with $\sin^2 \theta_W =0.235 $. 
Here, the first term in eq.(1.3) is important since the decay of 
the $Z^0$ boson into two photons is described by the parity violating part. 
In this case, the decay rate of the $Z^0 \rightarrow 2 \gamma $ can be described 
by the triangle diagrams which are basically the same as the $\pi^0$ decay process. 
The T-matrix of the $Z^0 \rightarrow 2 \gamma $ process has neither linear 
nor logarithmic divergence due to the Trace and parameter integrals, and therefore 
the total T-matrix of the $Z^0 \rightarrow 2 \gamma $ process is given as
$$ T_{Z^0 \rightarrow 2 \gamma} =-{g_z\over 6\pi^2} \left({2e\over 3}\right)^2 
(k_1^\alpha -k_2^\alpha) \varepsilon_{\mu \nu \rho \alpha} 
\epsilon_1^\mu \epsilon_2^\nu \epsilon_v^\rho \eqno{(1.4)}  $$
where ($k_1^\mu, \epsilon_1^\mu$) and ($k_2^\mu, \epsilon_2^\mu$) denote 
the four momentum and  polarization vector of two photons, respectively, and 
$\epsilon_v^\rho$ denotes the polarization vector of $Z^0$ boson. 
As we see below, this T-matrix should vanish to zero due to the symmetry nature 
of the two photon states (Landau-Yang theorem) \cite{landau,yang}. However, the present 
calculation rigorously proves that there is no room for the anomaly equation 
which was derived by regularizing the spurious linear divergence \cite{fujita,anomaly}.

\section{$\pi^0 \rightarrow \gamma+ \gamma $ process}
Before going to the discussion of the  $Z^0 \rightarrow 2 \gamma $ decay, 
we first review the calculation of the $\pi^0 \rightarrow 2 \gamma $ process 
which is first given by Nishijima 
\cite{nishi}. The interaction Lagrangian density ${\cal L}_I$ between fermion and pion 
can be given as eq.(1.2). In this case, the corresponding T-matrix 
for the $\pi^0 \rightarrow 2  \gamma $ reaction process can be written as
$$ T_{\pi^0 \rightarrow 2 \gamma} =  ie^2g_\pi \int {d^4p\over (2\pi)^4} {\rm Tr} 
\left[(\gamma \epsilon_1) {1\over p \llap/-M_N +i\varepsilon } 
(\gamma \epsilon_2) {1\over p \llap/-{k \llap/}_2-M_N +i\varepsilon }\gamma_5 
{1\over p \llap/+{k \llap/}_1-M_N+i\varepsilon  }  \right] $$
$$+ (1\leftrightarrow 2) .
   \eqno{(2.1)}   $$
where $\epsilon_1^\mu(\lambda_1)$ and $\epsilon_2^\mu(\lambda_2)$ denote the two 
polarization vectors of photons with the polarizations of $\lambda_1,\ \lambda_2$. 
In addition, $k_1^\mu, \ k_2^\mu$ denote the four momenta of two photons. 
Now, we can rewrite eq.(2.1) to evaluate the Trace parts as
$$ T_{\pi^0 \rightarrow 2 \gamma} =  2e^2g_\pi \int {d^4p\over (2\pi)^4} \ 
 {A_{\mu \nu} \epsilon_1^\mu \epsilon_2^\nu \over (p^2-M^2 )( (p-k_2)^2-M^2)
( (p+k_1)^2-M^2)   }  \eqno{(2.2)}    $$
where $A_{\mu \nu}$ is defined as
$$ A_{\mu \nu} \equiv {\rm Tr} [\gamma_\mu (p \llap/+M)\gamma_\nu 
(p \llap/-{k \llap/}_2+M) \gamma^5 (p \llap/+{k \llap/}_1+M)] .  \eqno{(2.3)} $$

\subsubsection{Linear Divergence Term}
Now the linear divergence term should correspond to the term which is 
proportional to $p^3$ in eq.(2.3), and thus we can show  
$$ A^{(3)}_{\mu \nu}={\rm Tr} [p \llap/ \gamma_\mu p \llap/ 
\gamma_\nu p \llap/ \gamma^5] =0   \eqno{(2.4)}  $$
which is due to the property of the Trace with $\gamma^5$ matrix. 

\subsubsection{Logarithmic Divergence Term}
Next, we should evaluate the $p^2$ term in eq.(2.3) which should correspond 
to the logarithmic divergence term. 
Now, this term can be written as
$$ A^{(2)}_{\mu \nu}={\rm Tr} [ \gamma_\mu p \llap/ \gamma_\nu p \llap/ \gamma^5] 
+{\rm Tr} [p \llap/ \gamma_\mu  \gamma_\nu p \llap/ \gamma^5] 
+{\rm Tr} [p \llap/ \gamma_\mu p \llap/ \gamma_\nu  \gamma^5] 
=0   \eqno{(2.5)} $$
and it also vanishes to zero by the Trace evaluation. 
Here we have made use of the following identity
$$ {\rm Tr} [ \gamma_\mu \  p \llap/ \  \gamma_\nu \  p \llap/ \gamma^5] 
=-4i\varepsilon_{\mu \rho \nu \sigma}  p^\rho  p^\sigma =0 . $$
Therefore, the T-matrix of the $\pi^0 \rightarrow 2  \gamma $ process has neither 
linear nor logarithmic divergences, and this is proved at the level of the Trace 
evaluation before the momentum integrations. 

\subsubsection{Finite Term}
Now, we can easily evaluate this momentum integral, and the result becomes 
$$  T_{\pi^0 \rightarrow 2 \gamma}  \simeq {e^2g_\pi\over 4\pi^2 M}
\varepsilon_{\mu \nu \alpha \beta}  k_1^\alpha k_2^\beta \epsilon_1^\mu 
\epsilon_2^\nu . \eqno{(2.6)} $$
As one sees, there is no divergence in this T-matrix calculation, and 
this is because the apparent linear and logarithmic divergences can be completely 
canceled out due to the Trace evaluation. In this respect, 
the corresponding T-matrix is finite and thus there is no chiral anomaly 
in this Feynman diagrams. This is, of course, well known, and the calculation of 
the T-matrix is explained quite in detail in the textbook 
of Nishijima in 1969  \cite{nishi}. \index{anomaly} 
\index{decay width of $\pi^0\rightarrow 2 \gamma$} 

\subsubsection{Decay Width of $\pi^0 \rightarrow 2\gamma$}
In this case, we can calculate the decay width $ \Gamma_{\pi^0 \rightarrow 2\gamma}$ as
$$ \Gamma_{\pi^0 \rightarrow 2\gamma}={1\over 8m_{\pi} |\bm{p}_1| |\bm{p}_2| 
 (2\pi)^2} \int \delta(m_{\pi}- |\bm{p}_1| - |\bm{p}_2| ) \delta( \bm{p}_1+ \bm{p}_2) |U|^2
d^3p_1d^3p_2   \eqno{(2.7)}  $$
where  $|U|^2$ is given as
$$ |U|^2={1\over 2} \sum_{\lambda_1, \lambda_2}  
|T_{\pi^0 \rightarrow 2 \gamma}  |^2    \eqno{(2.8)}  $$
where $\lambda_1$ and $\lambda_2$ denote the polarization state of two photons.  
The summation of the polarization state of two photons can be carried out 
by making use of the Coulomb gauge fixing which gives the polarization sum as 
$$ \sum_{\lambda=1 }^2  {\epsilon^*}^\mu_{\bm{k},\lambda}  \epsilon^\nu_{\bm{k},\lambda} 
= \left\{ 
\begin{array}{l}
 \left( \delta^{ab} -{k^a k^b\over \bm{k}^2} \right) 
\ \ \  \rm{for} \ \ \ \mu \not= 0,\nu \not= 0 \\ 
\ \ \\
 0 \ \ \ \hspace{2cm} \rm{for} \ \ \ \mu, \nu =0  .  \end{array} \right. 
 \eqno{(2.9)}  $$
After some calculations, we obtain
the decay width $ \Gamma_{\pi^0 \rightarrow 2 \gamma}$  as 
$$ \Gamma_{\pi^0 \rightarrow 2 \gamma} \simeq {\alpha^2\over 
16\pi^2}{g^2_\pi\over 4\pi}
{m_\pi^3\over M^2} \simeq 7.4 \ \ \ {\rm eV}  \eqno{(2.10)}  $$
which can be compared with the observed value  \cite{larin} 
$$  \Gamma_{\pi^0 \rightarrow 2 \gamma}^{exp} =7.8  \ \ {\rm eV} .  \eqno{(2.11)}  $$ 
As seen above, the calculation can well reproduce the observed data  of 
the life time of the $\pi^0\rightarrow 2 \gamma$ decay. 
Here, we take the value of ${g_\pi^2\over 4\pi} \simeq 8 $ which should be slightly 
smaller than the one determined from the nucleon-nucleon scattering experiments. 
This is clear since the $\pi^0\rightarrow 2 \gamma$ process should naturally include 
the effect of the nucleon form factor, in contrast to the value of the $\pi N N$ coupling 
constant obtained  from the nucleon-nucleon scattering experiments. In the case of 
$NN$ scattering, the nucleon form factors are introduced to accommodate the finite size 
effect of nucleons in the scattering process. 

Here, it should be important to note that the calculation of the decay width 
with the Coulomb gauge fixing is quite involved. On the other hand, the choice 
of the polarization sum of two photons
$$ \sum_{\lambda } {\epsilon^*}^\mu_{\bm{k},\lambda}  \epsilon^\nu_{\bm{k},\lambda}  
=-g^{\mu \nu}  \eqno{(2.11)}  $$
can also reproduce the correct decay width as given in eq.(2.9). In addition, 
the calculation with this choice of the polarization sum is much 
easier than the case with the correct expression of the polarization sum. 
However, the expression of eq.(2.11) cannot be justified for $\nu=\mu=0 $ case  
since the left hand side $\sum_{\lambda} 
| {\epsilon}^0_{\bm{k},\lambda} |^2 $ is always positive definite while 
the right hand side is negative. In this respect, the employment of 
eq.(2.11) is accidentally justified because of the special property 
of the amplitude in eq.(2.7).

\section{$Z^0 \rightarrow \gamma +\gamma $ process}
Now we can calculate the triangle Feynman diagrams which correspond to 
the $Z^0$ decay into two photons. The interaction Lagrangian density between 
$Z^0$ and fermions can be given in eq.(1.3), and therefore, the corresponding T-matrix 
for the triangle diagrams can be written as
$$ T_{Z^0 \rightarrow 2 \gamma} = g_z\sum_i e^2_i \int {d^4p\over (2\pi)^4} {\rm Tr} 
\left[ {1\over p \llap/-m_i +i\varepsilon } (\gamma \epsilon_1) 
 {1\over p \llap/-{k \llap/}_2-m_i +i\varepsilon } (\gamma \epsilon_2) 
{1\over p \llap/+{k \llap/}_1-m_i+i\varepsilon  }  (\gamma \epsilon_v)  \gamma_5 
\right] $$
$$ + (1\leftrightarrow 2)  \eqno{(3.1)} $$
where $m_i$ and $e_i$ denote the mass and charge of the corresponding fermions 
in the intermediate states. 

Here, we should make a comment on  Adler's calculation why he obtained the linear 
divergence in his paper. This is obviously connected to the fact that he did not 
include the second term $ (1\leftrightarrow 2) $ in his calculation of the T-matrix, 
and therefore he could not get rid of the linear divergence in his T-matrix 
evaluation \cite{adler}. This is not surprising at all since unphysical T-matrices 
may well have divergences which have nothing to do with nature.

\subsection{Decay Width with Intermediate Top Quark States}
Here, we take the top quark state as the intermediate state since it gives the largest 
contribution to the decay width. The evaluation of the T-matrix can be carried out 
in a straight forward way  just in the same manner as the $\pi^0 \rightarrow 2 
\gamma$ process. In order to avoid any confusions, we discuss the term by term 
in the integration of eq.(3.1). 

\subsubsection{Linear Divergence Term}
The leading term in the integration of eq.(3.1) at the large momentum of $p$ should have 
the following shape 
$$ T_{Z^0 \rightarrow 2 \gamma}^{(1)} \simeq  g_z e^2 \int {d^4p\over (2\pi)^4}\left[ { 
{\rm Tr} \left\{ p \llap/ \gamma_\mu  p \llap/ \gamma_\nu p \llap/
\gamma_\rho \gamma_5 \right\} \epsilon_1^\mu \epsilon_2^\nu \epsilon_v^\rho \over  
(p^2-s_0+i\epsilon )^3 } + (1\leftrightarrow 2) \right] . \eqno{(3.2)} $$
In this case, we can easily prove the following equation
$$ {\rm Tr} \left\{ p \llap/ \gamma_\mu  p \llap/ \gamma_\nu p \llap/
\gamma_\rho \gamma_5 \right\} =- {\rm Tr} \left\{ p \llap/ \gamma_\nu  p \llap/ 
\gamma_\mu p \llap/ \gamma_\rho \gamma_5 \right\} $$
and therefore we obtain
$$ T_{Z^0 \rightarrow 2 \gamma}^{(1)} \simeq  g_z e^2 \int {d^4p\over (2\pi)^4} { 
1 \over  (p^2-s_0+i\epsilon )^3 } \left[
{\rm Tr} \left\{ p \llap/ \gamma_\mu  p \llap/ \gamma_\nu p \llap/
\gamma_\rho \gamma_5 \right\} 
+{\rm Tr} \left\{ p \llap/ \gamma_\nu  p \llap/ \gamma_\mu p \llap/
\gamma_\rho \gamma_5 \right\}   \right] 
\epsilon_1^\mu \epsilon_2^\nu \epsilon_v^\rho =0 . \eqno{(3.3)} $$
In this respect, there is no linear divergence in the triangle diagrams and 
thus there is no need of the regularization. Therefore, this indicates that there 
exists no anomaly equation \cite{adler}. 

\subsubsection{Logarithmic Divergence Term}
The $p^2$ term of the numerator in eq.(3.1) contains the apparent logarithmic divergence.
However, we find that the logarithmic divergence term vanishes to zero 
in an exact fashion. First, we can calculate the Trace of the $\gamma-$ matrices 
and find the following shape for the logarithmic divergence term 
$ T_{Z^0 \rightarrow 2 \gamma}^{(0)}$ as 
$$ T_{Z^0 \rightarrow 2 \gamma}^{(0)} \simeq  g_z e^2 \int_0^1 dx\int_0^x dy 
\int {d^4p\over (2\pi)^4} { F(p,x,y) \over (p^2-s_0+i\epsilon )^3 }   \eqno{(3.4)} $$
where $F(p,x,y)$ is written  
$$ F(p,x,y)= {\rm Tr} \left\{ p \llap/ \gamma_\mu  p \llap/ \gamma_\nu a \llap/
\gamma_\rho \gamma_5 \right\} + {\rm Tr} \left\{ p \llap/ \gamma_\mu  b \llap/ 
\gamma_\nu p \llap/\gamma_\rho \gamma_5 \right\} +
{\rm Tr} \left\{ c \llap/ \gamma_\mu  p \llap/ \gamma_\nu p \llap/
\gamma_\rho \gamma_5 \right\}   \eqno{(3.5)} $$
where $a,b,c$ are given as
$$ a=-k_1(1-x)-k_2(1-y), \ \ \ b=-k_1(1-x)+k_2 y, \ \ \ c=-k_1 x+k_2 y  $$
After some tedious but straight forward calculation, we find that 
$$ T_{Z^0 \rightarrow 2 \gamma}^{(0)} =0  \eqno{(3.6)} $$
and therefore there is no need of the renormalization since the triangle diagrams 
are indeed all finite. 

\subsection{Finite Terms}
Here, one sees that the triangle diagrams with the axial vector coupling have neither 
linear nor logarithmic divergences. This is proved without any regularizations, and 
the total amplitude of $Z^0 \rightarrow 2 \gamma $ decay process is indeed finite. 
Here, we present the calculated decay width via top quarks since its contribution 
is the largest among all the other fermions. The finite term of the T-matrix can be 
written as 
$$ T_{Z^0 \rightarrow 2 \gamma} =  g_z e^2  \int_0^1 dx\int_0^x dy 
\int {d^4p\over (2\pi)^4} { A(x,y) \over (p^2-s_0+i\epsilon )^3 }   \eqno{(3.7)} $$
where $A(x,y)$ is given as 
$$ A(x,y)=-4im_t^2 (x+1-y) (k_1^\alpha -k_2^\alpha) \varepsilon_{\mu \nu \rho \alpha} 
\epsilon_1^\mu \epsilon_2^\nu \epsilon_v^\rho . $$
Here $m_t$ denotes the mass of the top quark. Therefore, the T-matrix becomes
$$ T_{Z^0 \rightarrow 2 \gamma} =-{g_z\over 6\pi^2} \left({2e\over 3}\right)^2 
(k_1^\alpha -k_2^\alpha) \varepsilon_{\mu \nu \rho \alpha} 
\epsilon_1^\mu \epsilon_2^\nu \epsilon_v^\rho \eqno{(3.8)}  $$

Now, we can prove that this should vanish to zero by choosing the system where 
$Z^0$ boson should be at rest. In this case, we can take the polarization vector 
$\epsilon_v^\rho$ as 
$$ \epsilon_v^\rho= (0, \bm{\epsilon}_v)  \eqno{(3.9)}  $$
which can satisfy the Lorentz condition of $k_\mu \epsilon_v^\mu =0 $. 
On the other hand, we can also choose the photon polarization vectors 
$\epsilon_1^\mu$ and  $\epsilon_1^\nu$ with the Coulomb gauge fixing as
$$ \epsilon_1^\mu=(0,\bm{\epsilon}_1) , \ \ \epsilon_1^\nu =(0,\bm{\epsilon}_2) 
\ \ \ \ \ {\rm with} \ \ \ \ \ \bm{k}\cdot \bm{\epsilon}_1=0, \ \ \  \bm{k}\cdot 
\bm{\epsilon}_2=0 . 
 \eqno{(3.10)}  $$ 
Further, we see that the $(k_1^\alpha -k_2^\alpha)$  should be expressed as 
$$ k_1^\alpha -k_2^\alpha =(0, 2\bm{k} ) .  \eqno{(3.11)}  $$
Therefore, we can easily prove by now that the T-matrix should be exactly zero 
due to the anti-symmetric nature of $ \varepsilon_{\mu \nu \rho \alpha}  $ where 
the non-zero part of the T-matrix should  always satisfy the condition that 
$\mu, \nu, \rho, \alpha$ should be different from each other. This zero decay 
rate is known as the Landau-Yang theorem \cite{landau,yang}

Here, we should make a comment on the branching ratio of 
$ {\Gamma_{Z^0\rightarrow 2\gamma}/ \Gamma}$, and  the present experimental 
upper limit shows  \cite{rmp}
$$ \left({\Gamma_{Z^0\rightarrow 2\gamma}/ \Gamma}\right)_{exp} 
 < 5.2 \times 10^{-5}   $$
which is consistent with zero decay rate. Therefore, the theoretical prediction 
of the branching ratio is indeed consistent with experiments.

\section{Conclusions}

We have presented a new calculation of the weak vector boson of $Z^0$ into two photons. 
The calculated T-matrix does not have either linear nor logarithmic divergences, 
and the finite term is shown to vanish to zero due to the symmetry nature 
of the two photon states. Therefore, there is no anomaly equation, and in fact, 
we confirm that the decay rate of the $\pi^0 \rightarrow 2 \gamma$ process is 
reproduced well by Nishijima's calculation. In addition, we see that all of 
the triangle diagrams do not have any divergences at all, and therefore theoretical 
scheme which involves the vacuum polarization is quite sound. In this respect, 
physical processess in connection with the self-energy of photon are all evaluated 
properly without employing the renormalization scheme, and thus there is no need of 
the renormalization of the photon self-energy. 
This is in contrast to the vertex correction which has still the logarithmic 
divergence at the present stage.

%\newpage


\vspace{2cm}

\begin{thebibliography}{99}


\bibitem{adler}
S.I. Adler, Phys. Rev. \textbf{177}, 2426 (1969)

\bibitem{nishi}
K. Nishijima, ``Fields and Particles " , (W.A. Benjamin,INC) 


\bibitem{bohr}
A. Bohr and  B.R. Mottelson, ``Nuclear Structure", 


\bibitem{mandl}
F. Mandl and G. Shaw,  ``Quantum Field Theory", 
(John Wiley \& Sons, 1993)


\bibitem{landau} 
L. D. Landau, Dokl. Akad. Nawk., USSR {\bf 60}, 207 (1948)

\bibitem{yang}
C. N. Yang, Phys. Rev. {\bf 77}, 242 (1950)



\bibitem{fujita}
T. Fujita, "Symmetry and Its Breaking in Quantum Field Theory", 
(2nd edition, Nova Science Publishers, 2011) 


\bibitem{anomaly}
T. Fujita and N. Kanda, J. Mod. Phys.  {\bf 3},  (2012) 665 




\bibitem{larin}
I. Larin et al. , Phys. Rev. Lett. {\bf 106}, 162303 (2011)


\bibitem{rmp}
C. Amsler et al. (Particle Data Group), Physics Letters {\bf B 667}, 1 (2008)


\end{thebibliography}
\end{document}